\documentclass{ws-procs10x7}
\newcommand{\gev}{{\mathrm GeV}}

\begin{document}

\title{Electroweak Measurements}

\author{Cristinel Diaconu}

\address{Centre de Physique des Particules de Marseille, IN2P3-CNRS, case 902, 163 Avenue de Luminy, Marseille, France  \\and Deutsches Elektronen Synchrotron, DESY, Notkestr. 85, 22607 Hamburg, Germany \\E-mail: diaconu@cppm.in2p3.fr}

\twocolumn[\maketitle\abstract{
The measurements of electroweak sector of the Standard Model are presented, including most recent results from LEP, Tevatron and HERA colliders. The robustness of the Standard Model is illustrated with the precision measurements, the electroweak fits and the comparisons to the results obtained from low energy experiments. The status of the measurements of the $W$ boson properties and rare production processes involving weak bosons at colliders is examined, together with the measurements of the electroweak parameters in $ep$ collisions. }]
\section{Introduction}
The Standard Model (SM) of the elementary particles has proven its robustness in the past decades due to extensive tests with increasing precision. 
In the present paper, the status of electroweak measurements in summer 2005 are presented. First, the precision measurements from LEP and SLD colliders will be summarised\footnote{The new averages of the $W$ boson and top quark masses, made public\cite{lepewwg} in july 2005 after the conference, are included in this paper together with the corresponding results of the electroweak fit.} and the confrontation with the low energy experiments will be reviewed.  Then production of weak bosons at LEP, Tevatron and HERA will be presented. The constraints obtained from an electroweak fit over DIS data will be described together with the latest measurements from polarized electron data at HERA. Finally, prospects for electroweak measurements at future colliders will be briefly reviewed.
\section{The experimental facilities}
 The LEP  collider stopped operation in 2000, after providing an integrated luminosity of more than $0.8$~fb$^{-1}$  accumulated by each of the four experiments (ALEPH, DELPHI, L3 and OPAL). The first stage (LEP I) running at $Z$ peak was continued with a second period (LEPII) with centre-of-mass energies up to 209, beyond the $W$ pair production threshold. The physics at the $Z$ peak was greatly inforced due to the polarised $e^+e^-$ collisions programme at the SLC. The SLD detector recorded a data sample corresponding to an integrated luminosity of 14~pb$^{-1}$, with a luminosity weighted electron beam polarisation of 74\%. 
\par The Tevatron $p\bar{p}$ collider completed a first period (Run I) in 1996. After an upgrade, including the improvement of the two detectors CDF and D0, a second high luminosity period started in 2002. In summer 2005 the delivered  integrated luminosity reached 1~fb$^{-1}$. The completion  of the second part of the programme (Run II) is forseen in  2009 with a goal of 4 to 8~fb$^{-1}$.  
 \par The unique $e^\pm p$ collider HERA is equiped with two detectors in collider mode H1 and ZEUS.  After a first period with unpolarised collisions (HERA I, 1993--2000), in the new stage (HERA II) the collider provides both electron and positron--proton collisions with $e^\pm$ beam polarisation of typically 40\%. The HERA programme will end in 2007 with a delivered luminosity around 700~pb$^{-1}$. 
 \par The situation of the high energy colliders in the last 15 years was therefore a favourable one, with  all three  combinations of colliding beams $e^+e^-$, $p\bar{p}$ and $ep$. The most precise testing of the weak interactions is done at $e^+e^-$ colliders (LEP and SLC), where $Z$ bosons are produced in the $s$--channel in a clean environment with sufficient luminosities. The hadronic collider (Tevatron) produces large samples of weak bosons and enables  complementary studies at higher energies, including the measurement of the top quark properties.  In $ep$ collisions, the exchange of space--like electroweak bosons  in the $t$ channel leads to new experimental tests of the Standard Model. The high energy experiments are complemented by low energy measurements that test the electroweak theory with high precision far below the weak boson masses. 
\section{The precision measurements, the electroweak fits and comparisons with low energy data}
\subsection{The precision measurements from high energy experiments}
\par
 The Standard Model is tested using a set of precison mesurements at $e^+e^-$ colliders close to the $Z$ peak. Those measurements, which were finalised recently\cite{z}, include data from the LEP experiments and the SLD detector at the SLC. 
\begin{figure}[ht]
\epsfxsize180pt
\figurebox{150pt}{190pt}{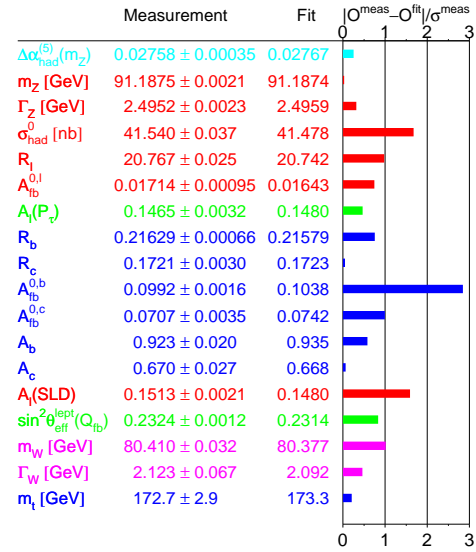}
\caption{The observables of the SM electroweak test compared with the values predicted from the fit. The pulls are also graphically shown and display  a consistent picture of the Standard Model. The largest deviation is found for  $A_{FB}^{b}$, slightly below $3\sigma$.}
\label{fig:ewpulls}
\end{figure}
\par
The two fermion production is measured using the flavour tag of the final state (leptons $\ell$ and the $b$ and the $c$ quarks).
More than 1000 measurements are used to extract a few observables that have simple relations to the fundamental parameters of the SM. The observables set include: the cross sections  and its dependence on the $\sqrt{s}$ (line shape given by $Z$ mass $m_Z$, width $\Gamma_Z$ and the hadronic pole cross section $\sigma^0_\mathrm{had}$) and on final state flavour (partial widths $R_{\ell,b,c}$), the  forward-backward asymmetries ($A_{FB}$), the left-right asymmetries ($A_{LR}$, measured for polarised beams or using the measured polarisation of the final state tau leptons). 
Moreover, the measured asymmetries are used in order to extract the asymmetry parameters that are directly related to the ratio of axial and vector couplings $A_f=2\frac{g^f_{V}/g^f_{A}}{1+(g^f_{V}/g^f_{A})^2}$. In the SM, $g^f_{V}/g^f_{A}=1-4|Q_f|\sin^2\theta_{\mathrm{eff}}^{\mathrm{f}}$, where $Q_f$ is the fermion charge and $\sin^2\theta_{\mathrm{eff}}$ the effective weak mixing angle, defined as the weak mixing angle including the radiative corrections.
\par
In order to relate the measurements to the fundamental constants of the SM, a simple parameter set is chosen. This includes the fine structure constant $\alpha(0)$, the strong coupling $\alpha_s$, the mass of the $Z$ boson $m_Z$ and the Fermi constant $G_F$ (related in practice to the $W$ boson mass). All fermion masses are neglected except the top quark mass $m_{\mathrm{top}}$. The Higgs boson mass $m_{\mathrm{Higgs}}$ plays a special role, due to its contribution to the radiative corrections. The observables are corrected for experimental effects and compared to the predictions from the Standard Model $O(\alpha,\alpha_s, m_Z, G_F, m_{\mathrm{top}}, m_{\mathrm{Higgs}})$. The QCD and electroweak radiative corrections are needed to match the experimental accuracy. The observables are therefore calculated with a precision beyond two loops. The fermion couplings $g^f_{A,V}$ that enter most of the observables depend via the radiative correction logarithmically on $m_{\mathrm{Higgs}}$ and quadratically on $m_{\mathrm{top}}$. This dependence allows the indirect determination of $m_{\mathrm{Higgs}}$ and $m_{\mathrm{top}}$.
\par
The running with energy of the electromagnetic coupling has to be taken into account in the radiative corrections. The running can be calculated analytically with high precision for the photon vacuum polarisation induced by the leptons and by the top quark. In constrast, the contribution due to the five light quark flavours $\Delta \alpha_{\rm had}^{(5)}$ is non-perturbative and has to be deduced from the measured $e^+e^- \rightarrow {\mathrm{hadrons}}$ cross section via the dispersion relations.  The determination of $\Delta \alpha_{\rm had}^{(5)}$ has been recently updated\cite{bolek} by including the new data from the $\rho$ resonance measured by CMD-2\cite{Akhmetshin:2003zn} and KLOE\cite{Aloisio:2004bu}. Despite a precision improvement by more than a factor of two in the $\rho$ region, the impact on $\Delta \alpha_{\rm had}^{(5)}$ precision is modest. QCD based assumptions may lead to an improved accuracy of the $\Delta \alpha_{\rm had}^{(5)}$ extraction\cite{deTroconiz:2004tr}. More precision measurements of hadron production in $e^+e^-$ collisions at low energy (in preparation) can bring significant improvements for the consistency checks of the SM. In addition, the hadronic vacuum polarisation estimates based on this data are also of high interest for the $(g-2)_\mu$ measurement, for which a $2.7\sigma$ discrepancy between the observation and the theory persists\cite{deTroconiz:2004tr,Bennett:2004pv}. 
\begin{figure}[ht]
\epsfxsize190pt
\figurebox{120pt}{160pt}{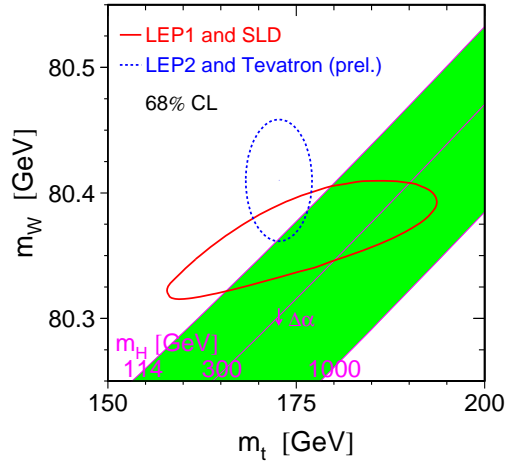}
\caption{The comparison of the direct measurements and indirect determinations of the top quark and W boson masses. The band indicates the SM constraint from the $G_F$ precise measurement for a range of Higgs masses. The confidence domain from the LEP1 and SLD data is compared with the direct measurement from LEP2 and Tevatron.  }
\label{fig:mtmw}
\end{figure}
\par
The list of the measured observables, using latest input from the LEP and SLC experiments\cite{lepewwg} is shown in figure~\ref{fig:ewpulls}. 
The fit of the observables in the SM framework is taken as a consistency check of the SM. The pulls of the observables plus the $\Delta\alpha^{(5)}_{\mathrm{had}}$ are also shown in figure~\ref{fig:ewpulls}. The picture displays  both the tremendous precision achieved by the electroweak tests and also the very good consistency of the SM. The most significant deviation, close to $3\sigma$, is given by the forward-backward asymmetry of the $b$--quarks, $A_\mathrm{FB}^b$. For this combined measurement, the individual values from various experiments and using different methods show very consistent results. 
\par Subtracting the visible partial widths $\Gamma_{\ell,b,c}$ deduced form the individual cross section from the total $Z$ width measured form the line shape, an invisible width can be deduced. Assuming that the invisible width is due to neutrinos with the same couplings as predicted by the SM, the number of neutrino flavours is determined to be $N_\nu=2.9840\pm0.0082$, in agreement with the SM expectation of three fermion generations.
\begin{figure}[ht]
 \epsfxsize190pt
 \figurebox{120pt}{160pt}{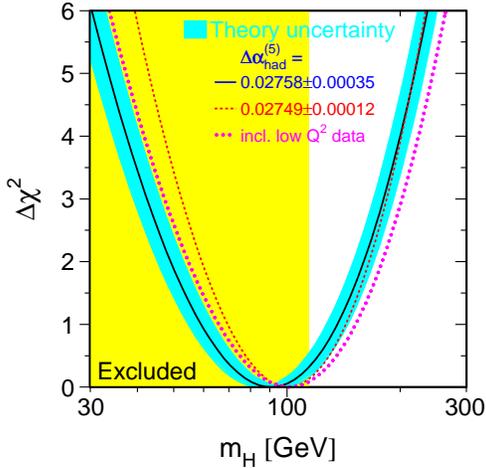}
 \caption{The $\chi^2$ of the SM fit, including the measured top quark mass, as a function of the Higgs boson mass. The results using two different estimations of  $\Delta \alpha_{\rm had}^{(5)}$ are shown together with the $\chi^2$ of the fit including the low energy data, described in the section~\ref{sec:low}. The theoretical error is indicated as a band.}
 \label{fig:blue}
 \end{figure}
\par 
An important ingredient for the SM consistency check is the direct measurement of the top quark mass. Together with the direct determination of the $W$ boson mass, to be discussed later, it consitutes a powerful test of the SM consistency. 
The new techniques and data samples from Run II  improved the measurement of the top quark mass at Tevatron. The new average\cite{Group:2005cc}, including recent measurement by the CDF and D0 collaborations is $m_\mathrm{top}=172.7\pm2.9$~GeV, a measureament which displays a dramatic improvement with respect to the previous error of 4.3~GeV (before the summer 2005). 
The comparison of the measured and fitted top and $W$ masses is shown in figure~\ref{fig:mtmw}.
The direct and indirect determinations are in agreement and favour low $m_\mathrm{Higgs}$.
Due to close connections between the electroweak correction involving the top quark and Higgs boson, the precison of the top mass measurement is crucial for the indirect constraints on the Higgs mass. The new fit of the Higgs boson mass from the electroweak model is shown in figure~\ref{fig:blue}. The new value of the fitted Higgs mass is $M_H=91^{+45}_{-32}$~GeV  with an upper limit within the SM formalism $M_H <186$~GeV~at 95\% CL. When the direct lower limit $M_\mathrm{Higgs}>114$~GeV is taken into account, the upper bound is found to be  $M_H <219$~GeV~at 95\% CL. 
\subsection{The electroweak precision measurements at lower energies}
\label{sec:low}
\begin{figure}[ht]
\epsfxsize190pt
\figurebox{120pt}{160pt}{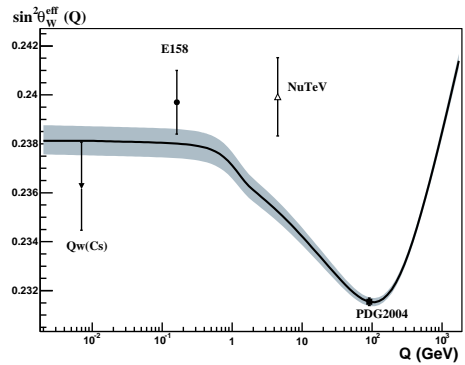}
\caption{The measurements of effective mixing angle at various energies ($Q$) compared with the theoretical prediction.}
\label{fig:runth}
\end{figure}
\par
Measurements of parity violation in the highly forbidden 6S--7S transition in Cs offer a way to test with high precision the SM\cite{cesium}. Due to its specific configuration with one valence electron above compact electronic shells, the theoretical calculation of the transition amplitude  achieves 0.5\% precision which allows the extraction with low ambiguities of the value for the nuclear weak charge\cite{Ginges:2003qt}. The weak charge of the nucleus depends linearly on the weak charges on the $u$ and $d$ quarks contained by the nucleons and interacting with the valence electron via $\gamma/Z$ in the t-channel. The weak charge $Q_w$ can therefore be written as a function of the effective weak mixing angle that can be measured in this way at very low (atomic-like) energies. 
\par 
The weak interactions can be tested at low energies via parity violating reactions. Using the end of beam at the SLC, polarized electrons are scattered off unpolarised atomic electrons (E158 expriment). The polarised  Moller scattering $e^-e^- \rightarrow e^-e^-$ offers the opportunity to extract $\sin\theta_\mathrm{eff}$  from cross section helicity asymmetry. This observable is related to the weak mixing angle that can be inferred with high precision\cite{Anthony:2005pm} at an energy of 160~MeV.
 \par
A classical way to access the weak sector at any energy is to measure neutrino induced processes. In the case of neutrino-nucleon scattering studied by the NuTeV experiment\cite{nutev}, the ratio of neutral current  to charged current cross sections $R_\nu=\sigma_\mathrm{CC}/\sigma_\mathrm{NC}$ is sensitive to the effective weak mixing angle, but subject to many systematic uncertainties related to the nucleon structure. Using both neutrino and anti-neutrino beams, the experimental results are combined  using the Pachos--Wolfenstein method $R_{-}=(\sigma^\nu_\mathrm{NC}-\sigma^{\bar{\nu}}_\mathrm{NC})/(\sigma^\nu_\mathrm{CC}-\sigma^{\bar{\nu}}_\mathrm{CC})$, for which large cancellations of systematical errors are expected. This ratio accesses the effective weak mixing angle and is also sensitive to the neutrino and quark weak couplings.
When SM couplings are assumed, the mixing angle measured by NuTeV is different from the SM prediction at 3.2$\sigma$ level. Missing pieces in either the theoretical prediction or in the theory error associated to the measurement are still under investigation\cite{Londergan:2005ua}.
\par
The measurement of the effective weak mixing angle at high and low energy can be used to test the electroweak running\cite{Czarnecki:2000ic}. The result is shown in figure~\ref{fig:runth}. Good agreement is found with the theoretical prediction, except for the NuTeV measurement discussed above. Precise measurements at energies beyond $M_Z$, as expected at the next $e^+e^-$ collider will test the predicted increase of $\sin^2 \theta^\mathrm{eff}_W$ with energy.
\subsection{The direct measurement of the running of  $\alpha$}
The running of the electromagnetic coupling has been observed by the OPAL experiment using low angle Bhabba scattering $e^+e^- \rightarrow e^+e^-$\cite{Abbiendi:2005rx}. The scattered electrons and positrons are detected close to the beampipe by two finely segmented calorimeters that allow the measurement of the scattering angles. The transfered momentum $t$ is therefore reconstructed and the variation of the cross section as a function of $t$ can be measured. 
 The cross section is directly proportional to the square of the electromagnetic coupling and inversely proportional to $t^2$. 
\begin{figure}[ht]
\epsfxsize190pt
\figurebox{120pt}{160pt}{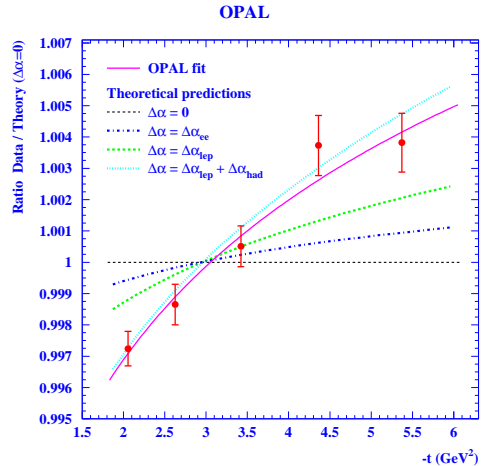}
\caption{$|t|$ spectrum normalized to the theoretical prediction for a fixed coupling ($\Delta\alpha=0$).}
\label{fig:runalpha}
\end{figure}
The electromagnetic coupling is expected to run with the collision scale, given by $t$.  The $t$ spectrum normalised to the theoretical prediction for a fixed coupling is shown in figure~\ref{fig:runalpha}. The difference of the measured event rates in $t$ bins and the theoretical prediction for no $\alpha$ running shows a clear dependence on $t$. This evidence at 5$\sigma$ level is compatible with the interpretation of $\alpha(t)$ running. When the pure electromagnetic running is taken into account in the theory, the remaining difference can be attributed to the hadronic component running that is in this way directly measured at 3$\sigma$ level.
\section{The weak bosons production and properties}
\subsection{The production of $W$ and $Z$ bosons}
The weak boson production mechanisms at LEP2 and Tevatron provide a test of the SM. In addition, the weak boson samples can be used to study their decay properties and further constrain the Standard Model. 
\par
$W$ pair production at LEP has been studied as a function of the centre-of-mass energy. The production cross-section variation with energy, 
flattens off at values around 15~pb, as expected from the SM, including the triple boson coupling $ZWW$. This behaviour is therefore directly related to the gauge structure of the Standard Model and constitutes an evidence for the non-abelian internal symmetry of the electroweak sector.
\par
\begin{figure}[ht]
\epsfxsize180pt
\figurebox{120pt}{160pt}{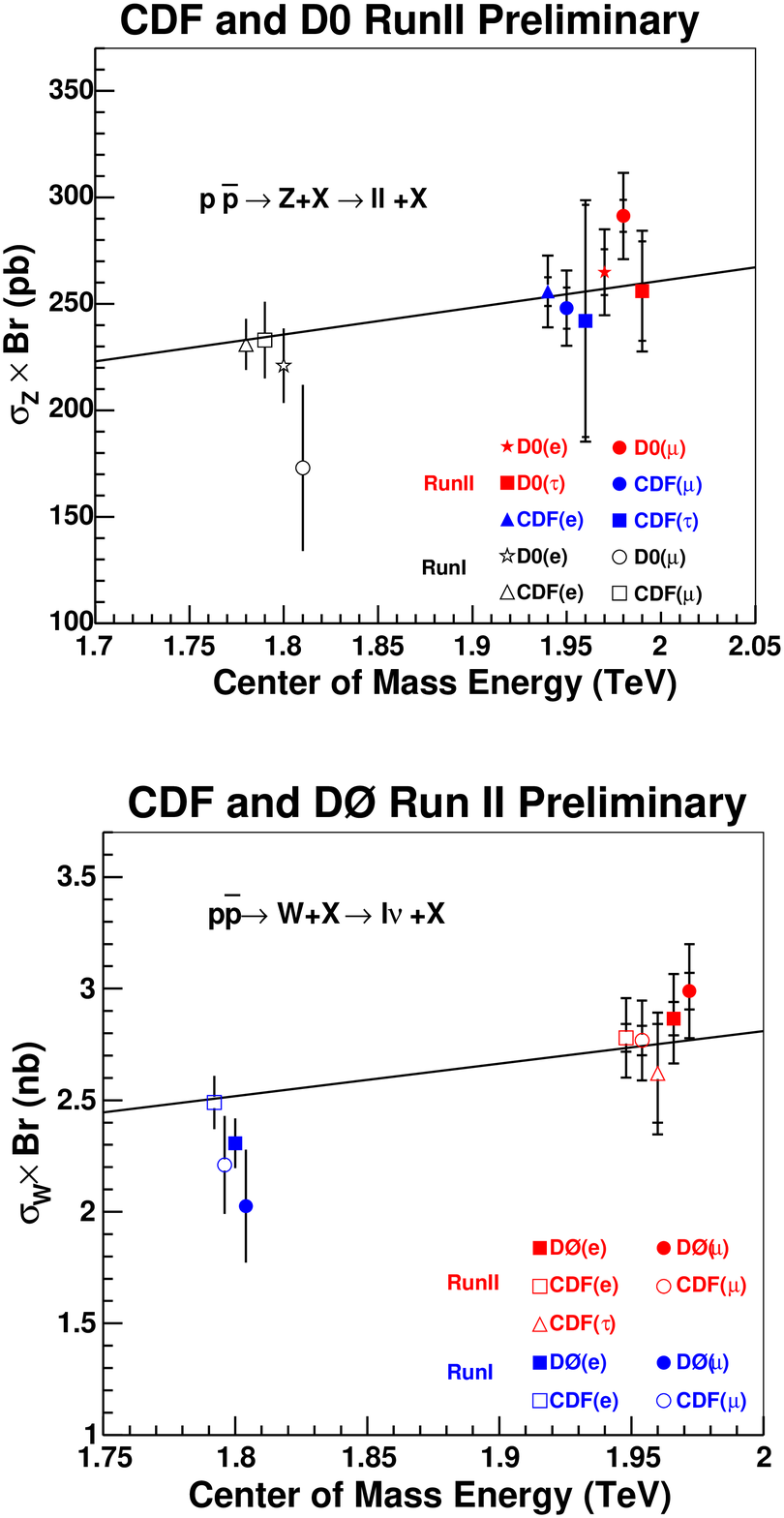}
\caption{Measurements of the $W$and $Z$ boson production cross section as a function of $\sqrt{s}$ at Tevatron. Data from CDF and D0 experiments obtained from various channels are compared with the theoretical prediction based on a NNLO QCD calculation.}
\label{fig:wztev}
\end{figure}

The $W$ and $Z$ bosons can be singly produced in $p\bar{p}$ collisions at Tevatron via the Drell-Yan process $q\bar{q}\rightarrow W$. The production crosss section is sensitive to the parton distribution functions. 
From this point of view, the production mechanism is also a convenient test ground for QCD, since the radiative corrections apply only to colliding partons and decouple from the produced bosons. The $W$ and $Z$ production cross sections measured in $p\bar{p}$ collisions at Tevatron are measured using the leptonic decay channels in $e$, $\mu$ or $\tau$. The results\cite{Acosta:2004uq} obtained from Run I ($\sqrt(s)=1.8$~TeV) and Run II ($\sqrt(s)=1.96$~TeV)  are shown  in figures~\ref{fig:wztev} and show a good agreement with the NNLO calculation\cite{Hamberg:1990np}. 
\subsection{The $W$ mass, width and branching ratios}
The $W$ harvest is also used to study the $W$ properties like the mass,the branching ratios and the width. The latest world average between the LEP and the Tevatron Run I measurements yields a $M_W=80.410\pm0.032$~GeV. The direct measurement agrees with the indirect determination from the LEP and SLD electroweak fit, including the $M_\mathrm{top}$ constraint $M_W=80.364\pm0.021$~GeV.
The  average  include a recent final result published by the OPAL Collaboration\cite{Abbiendi:2005eq}, for which a careful evaluation of the main systematical errors related to the colour reconnection and Bose--Einstein correlations together with an increased data sample allowed an improvement of the sytematical error from 70 to 56 MeV. This final precision of one LEP experiment is already better than the one of the combined result from Tevatron Run I and UA2 measurements\cite{Abazov:2003sv} $M_W^\mathrm{runI+UA2} = 80.456 \pm 0.059$~GeV.
\par
$W$ pair production at LEP is a favourable configuration to measure  $W$ branching fractions. The decay to electron channel  and to muon channel are found to be in very good agreement. However, the tau decay branching fraction is measured consistently by the four LEP experiments higher than the averaged electron and muon channel. This effect at 2.9$\sigma$ is for the time being one of the largest deviations in the SM precision tests. Future measurements of  $W\rightarrow \tau$ branching ratio are expected from Tevatron.
\par 
The $W$ width can be measured directly from the invariant mass spectrum at LEP, where high precision can be achieved via a kinematic fit based on the energy-momentum conservation. At Tevatron, where only the leptonic channel is measurable, the transverse mass spectrum is sensitive to the $W$ width in the tail at high mass. Finally, an indirect determination can be achieved exploiting the ratio of the $W$ and $Z$ cross section and using the precisely measured $Z$ parameters and the theoretical prediction of the cross section ratios, for which most of the QCD uncertainties cancel out. 
\par The present average of direct determination from LEP and Tevatron Run I is $\Gamma_W^\mathrm{direct}=2.123\pm0.067$, while the indirect determination from Run I data is $2.141\pm0.057$. A recent direct determination\cite{runIIw} from Run II data by D0 still display large errors $2.011\pm0.136$~GeV, while an indirect determination using the cross section ratio measured by CDF already improves the Run I value $2.079\pm0.041$~GeV. The value obtained from the LEP1 and SLD electroweak fit is extremely precise $2.091\pm0.002$~GeV and  in agreement with the direct and indirect determinations from LEP and Tevatron.
\subsection{The $A_\mathrm{FB}$ from $e^+e^-$ production at Tevatron.}
\begin{figure}[ht]
\epsfxsize180pt
\figurebox{120pt}{160pt}{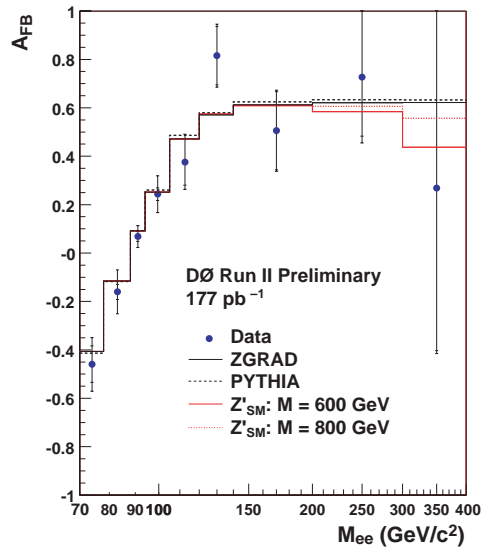}
\caption{The $A_\mathrm{FB}$ as a function of the $e^+e^-$ invariant mass measured at Tevatron.}
\label{fig:dyd0}
\end{figure}

The measurement of lepton pair production at Tevatron, produced via the Drell--Yan process $q\bar{q} \rightarrow \ell^+ \ell^-$ provides complex information about both the proton structure and the electroweak effects in new energy domain.  In particular, electron pair production can be used to measure the forward--backward asymmetry as a function of the pair mass. The result obtained by the D0 collaboration\cite{d0dy} is shown in figure~\ref{fig:dyd0}. The characteristic change of sign is observed around the $Z$ mass, similar to the much more precise measurement from LEP. From the measured asymmetry, the effective weak mixing angle is extracted by CDF\cite{Acosta:2004wq} $\sin^2\theta^\mathrm{eff}_W = 0.2238 \pm 0.0050$, in good agreement with the value measured at LEP from the forward--backward asymmetry  $0.2324\pm0.0012$.
At large invariant masses, the deviation from the SM prediction may indicate the production of a heavier neutral boson $Z'$, in case it has similar couplings to fermions as in the SM.
\subsection{Rare $W$ and $Z$ production processes}
At LEP2, in contrast to $W$ pair production, single boson production ($W$ or $Z$) is a rare process with cross sections below 1~pb.
The final state contains four fermions, with only one fermion pair consistent with the boson mass. The comparison to the Standard Model provides a test in a low density phase space region, where new phenomena can occur. The cross section is typically 0.6-0.9~pb for single $W$ production and  0.5--0.6~pb for single $Z$ production at $\sqrt{s}=182-209$~GeV.
\begin{table*}[ht]
\caption{Summary of the results of searches for events with isolated
leptons, missing transverse momentum and large hadronic transverse momentum $p_T^X$ at HERA. The number
of observed events is compared to the SM prediction. The $W^\pm$
component is given in parentheses in percent. The statistical and systematic uncertainties added
in quadrature are also indicated.}
\label{tab:heraisolep}
 \renewcommand{\arraystretch}{1.4}
 \footnotesize
\begin{center}
\begin{tabular}{|c|c|c|c|c|} \hline
  \multicolumn{2}{|c|}{ obs./exp.(W) } & Electron & Muon & Tau$^{H1:105~pb^{-1}}$ \\\hline
{\large \bf H1}                & Full sample &  25 / $20.4~\pm2.9~(68\%)$   & 9 / $5.4~\pm1.1~(82\%)$  & 5 / $5.8~\pm1.4~(15\%)$ \\ \cline{2-5}
{\footnotesize $211$~pb$^{-1}$} & $p_T^X~>~25\gev$ &  11 / $3.2~\pm0.6~(77\%)$   & 6 / $3.2~\pm0.5~(81\%)$  & 0 / $0.5~\pm0.1~(49\%)$ \\ 
\hline
\hline
 {\large \bf ZEUS }   & Full sample &  24 / $20.6~^{+1.7}_{-4.6}~(17\%)$   & 12 / 11.9~$^{+0.6}_{-0.7}$ (16\%)  & 3 / 0.40~$^{+0.12}_{-0.13}$~(49\%) \\  \cline{2-5}
{\footnotesize $130$~pb$^{-1}$} & $p_T^X~>~25\gev$ &  2 / $2.90~^{+0.59}_{-0.32}~(45\%)$   & 5 / 2.75~$^{+0.21}_{-0.21}$ (50\%)  & 2 / 0.20~$^{+0.05}_{-0.05}$~(49\%) \\  
\cline{2-5}
{\footnotesize $106$~pb$^{-1}_{new}$} & $p_T^X~>~25\gev$ &  1 / $1.5~\pm0.2~(78\%)$   &   &  \\  
 \hline
\end{tabular}
\end{center}
\end{table*}
\par
At Tevatron, where weak bosons are massively singly produced, boson pair occur with a much lower rate. The associated $W\gamma$ or $Z\gamma$ production processes, with the weak bosons decaying into leptons,  have cross sections close to 20~pb and 5~pb respectively\cite{Acosta:2004it}. The pair production of weak bosons is a particularly interesting process due to the spectacular final state and because it constitutes the main background for the search of the Higgs boson for $m_\mathrm{Higgs}>150$~GeV. While $WW$ production has been measured\cite{Acosta:2005mu,Abazov:2004kc}, the search for $WZ$ and $ZZ$ production\cite{Abazov:2005ys,Acosta:2005pq} have not been successful with the present luminosity and upper limits around $13-15$~pb at 95\% CL have been calculated, for a total SM prediction of 5~pb.
\begin{figure}[ht]
\epsfxsize190pt
\figurebox{120pt}{160pt}{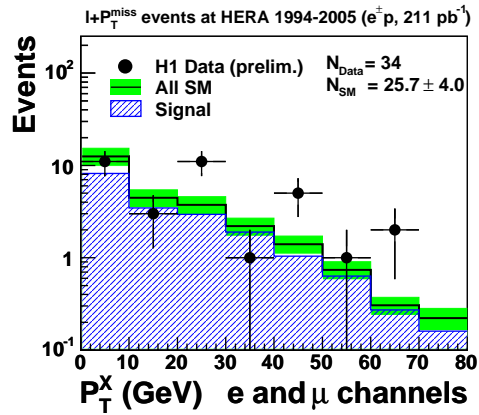}
\caption{The transverse momentum of the hadronic system in events with isolated electrons or muons  and missing $P_T$ measured by the H1 experiment at HERA. }
\label{fig:isolptx}
\end{figure}
\par
The single $W$ can also be produced in $e^\pm p$ collisions at HERA, with a cross section around 1~pb. The main production mechanism involves a fluctuation of photon emitted by the electron into a hadronic state, followed by the collision with the proton which leads to a $qq'$ fusion into a $W$ boson. In case of leptonic decay of the $W$, the final state consist of an high transverse momentum isolated lepton, missing transverse energy and possibly a low $P_T$ hadronic system $X$. The H1 collaboration reported\cite{Adloff:1998aw,Andreev:2003pm} the observation of such events and measured the cross section as a function of the hadrons transverse momentum ($P_T^X$). While a good agreement is observed at low $P_T^X$, a few spectacular candidates are observed at large $P_T^X$. The events continue to be observed at HERA II\cite{h1isol} by the H1 Collaboration which has analysed a total sample corresponding to an integrated luminosity of 211 pb$^{-1}$. The distribution of the transverse momentum of the hadronic system is shown in figure~\ref{fig:isolptx}, where the excess of observed events at large $P_T^X$ is visible. 
\par
The ZEUS collaboration has investigated their data with a different analysis strategy, with less purity for the SM $W$ signal, the full HERA I data set and observes some events at large $P_T^X$, but no prominent excess above the SM prediction\cite{Chekanov:2003yt}. Recent modified analysis of the electron channel only, performed using a similar amount of data (but combining partial HERA I and II data sets) also do not support the H1 observation\cite{zeusisolnew}. ZEUS collaboration observes events with tau leptons, missing transverse momentum and large hadronic transverse momentum, while no such event is observed by H1. The results are summarized in table~\ref{tab:heraisolep}. 
More incoming data will help to clarify this issue, which is at present one of the most intriguing results from HERA.
\section{The measurement of the electroweak effects at HERA}
\subsection{Combined QCD/Electroweak fit of DIS data}
The deep inelastic collisions at HERA are classically used to extract the proton structure information\cite{Adloff:2003uh,Chekanov:2002pv}. More than 600 measurement points of the charged and neutral current double differential cross section $d\sigma^\mathrm{CC,NC}/dx dQ^2$,  where $x$ is the proton momentum fraction carried by the struck quark and $Q^2$ is the boson virtuality, have been used together with other (fixed target) measurements to extract the parton distribution functions. Due to the high $ep$ centre--of--mass energy (320~GeV), the proton is investigated down to scales of $10^{-18}$m. The point--like nature of quarks is  tested in the electroweak regime, where the  proton is "flashed" with weak bosons. This experimental configuration allows to separate quark flavours within the proton and to improve the precision with which the parton distribution functions are extracted.  Conversely, the electroweak sector can be investigated using the knowledge of the proton structure. 
\begin{figure}
\epsfxsize190pt
\figurebox{120pt}{160pt}{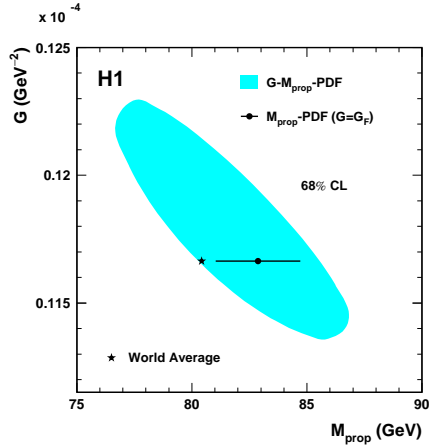}
\caption{The~allowed~region~at~65\%~CL~in the plane  $(G_F,M_W^\mathrm{prop})$ obtained from the combined electroweak--QCD fit of the DIS data.}
\label{fig:mprop}
\end{figure}
\par
Recently, a consistent approach has been adopted by the H1 Collaboration\cite{Aktas:2005iv}, performing a combined QCD--electroweak fit.  The strategy is to leave free in the fit the EW parameters together with the  parameterisation of the parton distribution functions.  
\par
An interesting result is related to the so--called propagator mass $M^\mathrm{prop}_W$, that enters a model independent parameterisation of the CC cross section: 
$${ \displaystyle
{\frac{d^2\sigma^{{\pm}}_{\rm CC}}{dxdQ^2}}=
\frac{G^2_F}{2\pi x}\left(\frac{M^2_W}{M^2_W+Q^2}\right)^2 
\tilde{\Phi}_{CC}},
$$ where $G_F$ is the Fermi constant and $\tilde{\Phi}_{CC}$ is the reduced cross section that encapsulates the proton structure in terms of parton distribution functions. 
If the Fermi constant $G_F$ and the propagator mass are left free in the fit,
an allowed region in the $(G_F,M_W^\mathrm{prop})$ plane can be measured. The result is shown in figure~\ref{fig:mprop}. By fixing $G_F$ to the very precise experimental measurement, the propagator mass can be extracted and amounts in this analysis to $M^\mathrm{prop}_W=82.87 \pm 1.82 {(\mathrm{exp.})} {}^{+0.30}_{-0.16} {(\mathrm{model})} \mbox{ GeV}$, in agreement with the direct measurements. 
\par
If the framework of SM model is assumed, the $W$ mass can be considered as a parameter constrained by the SM relations and entering both the cross section and the higher order correction. In this fitting scheme, where $M_W$ depends on the top and Higgs masses, the obtained value from DIS is $M_W=80.709 \pm 0.205 (\mathrm{exp}) {}^{+0.048}_{-0.029}  (\mathrm{mod})\pm 0.025  (\mathrm{top})  \pm 0.033(\mathrm{th}) -0.084 (\mathrm{Higgs}) \mbox
{ GeV} $, in good agreement with other  indirect determinations and with the world average. The fit value can be converted into an indirect $\sin\theta_W$ determination using the relation $ \sin^2\theta_W = 1 - \frac{M^2_W}{M^2_Z}$, assumed in the on mass shell scheme. The  result $\sin^2 \theta_W=0.2151\pm 0.0040_{\mathrm exp.} {}^{+0.0019}_{-0.0011}|_{th}$, obtained for the first time in from $e^\pm p$ collisions,  is in good agreement with the value of $0.2228\pm0.0003$ obtained from the measurements in $e^+e^-$ collisions at LEP and SLC. 
\begin{figure}
\epsfxsize190pt
\figurebox{120pt}{160pt}{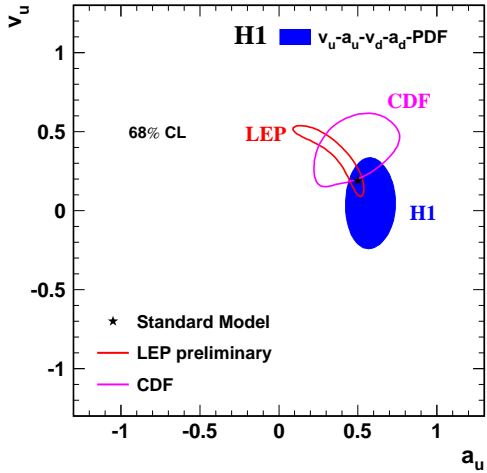}
\caption{Axial and vector couplings of the $u$--quark 
measured from the combined electroweak--QCD fit at HERA and compared with measurements from LEP and Tevatron.}
\label{fig:auvu}
\end{figure}
\par 
Due to the $t$-channel electron-quark scattering via $Z$ bosons, the DIS cross sections at high $Q^2$ are sensitive to light quark axial ($a_q$) and vector ($v_q$) coupling to the $Z$. This dependence includes linear terms with significant weight in the cross section which allow to determin not only the  value but also the sign of the couplings. In contrast, the measurements at the $Z$ resonance (LEP1 and SLD) only access $av$ or $a^2+v^2$ combinations. Therefore there is an ambiguity between axial and vector couplings and only the relative sign can be determined. In addition, since the flavour separation for light quarks cannot be achieved experimentally, flavour universality assumptions have to be made. The Tevatron measurement \cite{Acosta:2004wq} of the Drell-Yan process allows to access the couplings at an energy beyond the $Z$ mass resonance, where linear contributions are significant. The measurements of the $u$--quark couplings obtained at HERA, LEP and Tevatron are shown in figure~\ref{fig:auvu}. 
The data to be collected at Tevatron and HERA as well as the use of polarized $e^\pm$ beams at HERA open interesting oportunities for the light quarks couplings measurements in the near future.
\subsection{$e^\pm$ collision with polarised lepton beam}
\begin{figure}
\epsfxsize190pt
\figurebox{120pt}{160pt}{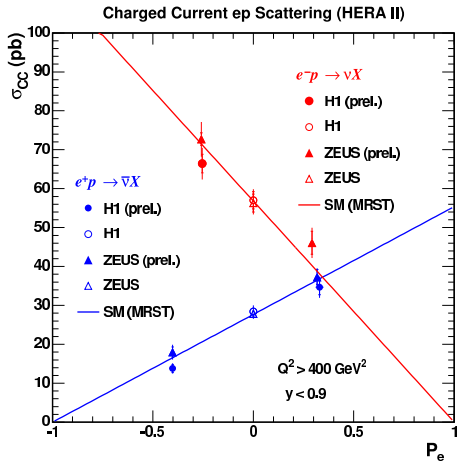}
\caption{The dependence of the total CC cross section of the $e^\pm$-beam polarization at HERA.}
\label{fig:ccpol}
\end{figure}

The polarisation of the electron beam at HERA II allows a test of the parity non-conservation effects typical for the electroweak sector. 
The most prominent effect is predicted in the CC process, for which the cross section depends linearly on the $e^\pm$--beam polarisation: $\mathrm{\sigma^{e^{{\pm}} p} (P) =(1 {\pm} P) \sigma^{e^{{\pm}} p}_{P=0}}$. The results\cite{ccpol,ccpolz} obtained for the first time in $e^{\pm}p$ collisions are shown in figure~\ref{fig:ccpol}. The expected linear dependence is confirmed and constitute supporting evidence for the V-A structure of charged currents in the Standard Model, a property already verified more than 25 years ago, by measuring the polarisation of positive muons produced from $\nu_\mu$--Fe scattering by the CHARM experiment\cite{Jonker:1979md}.
\par 
\section{Outlook}
The present experimental activity towards electroweak measurements continues to provide increasing endurance tests for the Standard Model. The LEP analyses are  final in many aspects and the results still play a key role in the present understanding of the electroweak symmetry breaking mechanism. The incoming data from Tevatron has good chance to take over in many aspects, especially concerning the weak boson properties, but also to extend the area of the measurements beyond LEP energies. At HERA, a consistent approach of the electroweak and QCD processes will certainly bring valuable information in the near future. The low energy measurements provide not only a cross check but also a solid testing ground for the electroweak sector, for which surprises are not excluded.
\par
The future colliders will test  the electroweak sector with high precision\cite{ewfut}.
At the LHC, the electroweak physics will mainly profit from the huge increase in the weak boson production cross-section. In the foreseen experimental condition the precision on the W mass measurement should approach 15 MeV while  the  top quark mass will be measured at 1 GeV level. The next $e^+e^-$ linear collider will improve the precision on $M_W$ to below 10 MeV while the top mass will be measured to 100~MeV. Similarly to the present situation, the precise measurements of the electroweak sector will allow to set indirect limits on the new physics, that might well be beyond the direct reach of the future colliders.
\section*{Acknowledgments}
I would like to thank the following people for kind assistance during the preparation of this contribution: Max Klein, Martin Grunewald, Pippa Wells,
Dmitri Denisov, Jan Timmermans, Bolek
Pietrzyk, Emmanuelle Perez, Matthew
Wing, Richard Hawkings, David Waters,
Chris Hays and Dave South.

\end{document}